\documentclass[pra,aps,twocolumn]{revtex4}
\usepackage{amsmath}
\usepackage{amsfonts}
\usepackage{amssymb}
\usepackage{color}
\usepackage{graphicx}

\begin{document}

\title{Strong non-equilibrium effects in spin torque systems}

\author{Tim \surname{Ludwig}$^1$, Igor S. \surname{Burmistrov}$^{2,3}$, Yuval \surname{Gefen}$^4$, 
Alexander \surname{Shnirman}$^1$}
\affiliation{$^1$Institut f\"ur Theorie der Kondensierten Materie, Karlsruhe Institute of Technology, D-76128 Karlsruhe, Germany}
\affiliation{$^2$L.D. Landau Institute for Theoretical Physics RAS, Kosygina street 2, 119334 Moscow, Russia}
\affiliation{$^3$Laboratory for Condensed Matter Physics, National Research University Higher School of Economics, 101000  Moscow, Russia} 
\affiliation{$^4$Department of Condensed Matter Physics, Weizmann Institute of Science, 76100 Rehovot, Israel}

\begin{abstract}

We consider a problem of persistent magnetization precession in a single domain ferromagnetic nano particle under 
the driving by the spin-transfer torque. We find that the adjustment of the electronic distribution function in the particle renders this state unstable. Instead, abrupt switching of the spin orientation is predicted upon increase of the spin-transfer torque current. On the technical level, we derive an effective action of the type of Ambegaokar-Eckern-Sch\"{o}n 
action for the coupled dynamics of magnetization (gauge group $SU(2)$) and voltage (gauge group $U(1)$).

\end{abstract}

\maketitle

\section{Introduction}

The dynamics of magnetization of single-domain ferromagnets (e.g. nanoparticles, quantum dots) 
under the influence of spin-transfer torque (STT)~\cite{Slonczewski1996L1,BergerSpinTorque,RevModPhys.77.1375} is a 
cornerstone of spintronics. Arguably, the most intriguing regime of behavior is that of persistent 
precession of magnetization driven by STT. This effect is predicted even in ferromagnetic nano 
particles with no actual anisotropy, i.e., the only preferred direction is due to the applied field.   
Here we revisit the semiclassical magnetization dynamics of such systems and find that this regime 
may become unstable owing to the non-equilibrium distribution on the dot.
Consequently, the persistent precession is replaced by a different steady state spin dynamics, 
but can possibly be restored by further introducing internal energy and spin equilibration processes in the 
quantum dot. 

Spin dynamics in STT driven systems is known to be very rich. Two main 
effects, i.e. persistent precession and stochastic switching, should be distinguished. In systems with strong magnetic 
anisotropy and, thus, multiple local stability at equilibrium, the STT can facilitate stochastic switching between the stable 
solutions, see, e.g., Refs.~\cite{ApalkovVisscher,SwiebodzinskiPRB}. For example in an easy-axis situation the 
switching between the north pole and the south pole of the Bloch sphere can be enhanced by 
inducing tunneling spin current between a pinned ferromagnet and a nanoparticle with freely rotating 
magnetization (such an arrangement is usually referred to as magnetic tunnel junction). 
The theoretical method of choice in this context is the stochastic Landau-Lifshitz-Gilbert (LLG) equation with 
an additional (Slonczewski's) term~\cite{Slonczewski1996L1,BergerSpinTorque} due to the STT and 
with corresponding Langevin sources.
 
The other effect, i.e., the persistent precession, is the main focus of the present analysis.
A strong enough STT current can shift an equilibrium stable solution, e.g. the 
magnetization directed along the external magnetic field, into a different, dynamically stable stationary solution. 
The latter may be characterized, e.g., by persistent precession of the magnetization around the direction of the 
external field (or, more generally, around the effective field determined by the anisotropy). Persistent precession states were predicted theoretically back in 1996~\cite{Slonczewski1996L1,BergerSpinTorque}, and have been subsequently 
observed~\cite{KiselevNature2003}. Since then the spin torque nano-oscillators are being 
actively investigated (see e.g., Refs.~\cite{Zeng2013,Choi2014,Ramaswamy2016}).
Rotating magnetization may, in turn, induce spin and charge currents across the magnetic tunnel junction~\cite{BergerInverse,PRBTserkovnyak2008}.

Usually the two phenomena (spin current inducing magnetization 
precession and the latter inducing spin and charge currents) are treated separately, although they are obviously 
intimately related. Here, extending Ref.~\cite{PRLShnirman2015}, we derive an effective action of the Ambegaokar-Eckern-Sch\"on (AES) type~\cite{AES_PRL,AES_PRB}, which governs the dynamics of both the charge ($U(1)$) and the spin ($SU(2)$) 
degrees of freedom. This allows us to obtain equations of motion describing simultaneously the induction of spin precession by current and the generation of current by rotating magnetization.

Our main result concerns the effect of the non-equilibrium distribution function of electrons in the nanoparticle
generated by the applied voltage and by the ensuing spin precession. Here it is important to 
understand the major difference between, e.g., molecular systems with large spin and the itinerant ferromagnetic nanoparticles. In the former case the dynamical degrees of freedom are restricted to the spin itself, whereas 
the dissipation and the driving are provided by macroscopic electronic reservoirs 
with fixed distribution functions, which are coupled to the spin. These reservoirs may be formally integrated out 
rendering an effective dissipative action for the spin alone. By contrast, in an itinerant ferromagnetic nanoparticle, 
the internal microscopic state of the electrons described by the distribution function plays a decisive role. 
Formally, this distribution function is completely enslaved to the trajectory in time of magnetization 
and voltage as well as to the boundary conditions in the leads. However, it may be driven far from equilibrium, which 
in turn influences strongly the effective action of the magnetization and voltage
(see the discussion in Chapter 11.6 of Ref.~\onlinecite{AltlandSimonsBook}). This has been indeed
realized in Ref.~\cite{PRBBaskoVavilov2009}, where the magnetization dynamics was 
assumed to be slow compared to the relaxation time of the distribution function -- the latter, then, deviated only 
slightly from instantaneous equilibrium. In this paper we focus on the 
opposite regime of rather fast magnetization dynamics and high voltage leading to 
strong non-equilibrium conditions (cf.~Ref.~\onlinecite{PhysRevB.81.024416}). Note that earlier treatments of magnetic tunnel junctions overlooked frequently the role of the sink, which drains the current flowing into the dot from the fixed ferromagnet. We will show that proper treatment of the sink is crucial for the correct description of the dynamics.

Another possibility to circumvent the non-equilibrium distribution function is 
to assume strong energy and spin relaxation inside the dot~\cite{ChudnovskiyPRL}. The distribution function 
in the dot is, then, Fermi-Dirac, and the effective action for the magnetization derived in Ref.~\cite{ChudnovskiyPRL} 
results in LLG equations with Slonczewski's spin-transfer torque term. This, in turn, results in persistent precession of magnetization and in interesting fluctuation relations for statistics of charge, spin, and heat. The latter were studied recently in Ref.~\cite{VirtanenHeikkila2016}, using the effective action of the AES type derived in Ref.~\cite{PRLShnirman2015}.

Here we analyze the opposite situation. We assume that the only relaxation in the dot is due to the tunneling to the leads. In this case, we obtain the LLG equations again. Yet, the spin and the charge currents appearing in these equations are strongly modified by the adjustment of the electron distribution function to the state of persistent precession. 
This renders the state of persistent precession unstable. Upon increasing the applied voltage, the unique stable solution jumps abruptly from the magnetization oriented towards the north pole of the Bloch sphere to its south pole. Neither orientation involves spin precession. We note in passing that our formalism reproduces the persistent precession solution of Ref.~\cite{ChudnovskiyPRL} as well as the precession generated current of Ref.~\cite{PRBTserkovnyak2008} once we enforce the equilibrium electron distribution in the dot (nanoparticle).

The paper is organized as follows. In Sec.~\ref{sec:System} we describe the system under investigation.
Sec.~\ref{sec:Derivation} is devoted to the derivation of the effective action. In Sec.~\ref{sec:Solution} we 
derive the quasi-classical equations of motion, taking into account the fact that the distribution function of electrons
in the dot is driven far from equilibrium. We find the stationary regimes of the driven system and analyze their stability. 
Finally, in Sec.~\ref{sec:Discussion} we discuss the obtained results.

\section{The system}
\label{sec:System}
\begin{figure*}
\begin{center}
\includegraphics[width=0.5\textwidth]{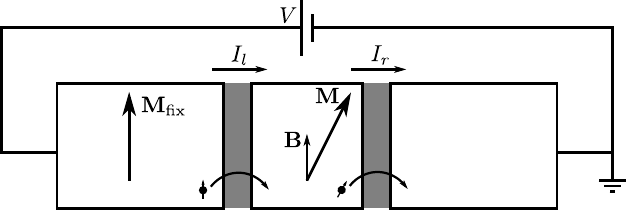}
\end{center}
\caption{Schematic view of the system: A ferromagnetic quantum dot (center) is exposed to an external magnetic field and tunnel coupled to two leads. The left lead is a ferromagnet with a fixed magnetization direction. The right lead is non-magnetic. A non-equilibrium situation is generated by a voltage $V$ that is applied across the system.} \label{System}
\end{figure*}
We consider a quantum dot tunnel coupled to two leads, one magnetic lead and one non-magnetic lead, see Fig. \ref{System}. The Hamiltonian for the full system is 
\begin{equation}
H=H_{dot} + H_{l} + H_{r}  + H_{tun.}\ .
\end{equation} 
The left lead is chosen to be the ferromagnetic one with fixed magnetization and therefore with fixed exchange field $M_\mathrm{fix}$ along the $z$-axis~\footnote{Note that for the present calculation only the DOS at the Fermi-surface is important. The actual direction of the fixed magnetization might however enter in the renormalization of the magnetic field (see discussion after Eq.~(\ref{Vsigma})).}. We also include an electrostatic potential~\footnote{To obtain the physical units, we have to replace $V \rightarrow e V$, $V_d \rightarrow e V_d$ and $\mathbf B \rightarrow \lambda \mathbf{B}$, $\mathbf M \rightarrow \lambda \mathbf M$, $\mathbf M_\mathrm{fix} \rightarrow \lambda \mathbf M_\mathrm{fix}$, $\mathbf B_\mathrm{exc} \rightarrow \lambda \mathbf{B}_\mathrm{exc}$, where $\lambda \equiv \hbar g e/(2mc)$ and $g\approx 2$; note that $e<0$ and, thus, $\lambda <0$.} on the left lead, $V$, i.e.,
 \begin{equation}
 H_{l}= \sum_{n=1}^{N_l}\sum_\sigma\int\frac{dk}{2\pi} \left(\epsilon^{{\phantom \dagger}}_{nk} -  \frac{M_\mathrm{fix}}{2} \sigma + V\right) c^\dagger_{nk, \sigma} c^{\phantom \dagger}_{nk, \sigma}\ .
\end{equation} 
Here $n$ counts the channels of the left lead 
$n=\{1,\dots,N_l\}$, whereas $k$ denotes the wave number in the channel.
The right lead is assumed to be grounded and non-magnetic, i.e.,
\begin{equation}
H_{r}= \sum_{n=N_l+1}^{N_l + N_r}\sum_\sigma \int\frac{dk}{2\pi} \,\epsilon^{\phantom \dagger}_{nk}\, 
c^\dagger_{nk,  \sigma} c^{\phantom\dagger}_{nk,  \sigma}\ .
\end{equation} 
Here $n\in \{N_l+1,\dots,N_r\}$ counts the channels of the right lead.
We employ the universal Hamiltonian for the dot~\cite{PRBKurland2000} ignoring the Cooper channel, i.e.
\begin{eqnarray}\label{eq:Hdot}
&&H_{dot}   =  H_{dot,0}  - \mathbf{B} \mathbf{S} -J \mathbf{S}^2 +  E_{c} (N-N_0)^2\ , \\
&&H_{dot,0}  =  \sum_{\alpha, \sigma} \epsilon^{\phantom \dagger}_\alpha a^\dagger_{\alpha, \sigma} a^{\phantom \dagger}_{\alpha, \sigma}\ ,
\end{eqnarray}
where $J$ is the exchange constant, $E_{c}$ is the charging energy, and $N_0$ represents the background charge. 
The spin operator is given by
\begin{equation}
\mathbf{S}= \frac{1}{2} \sum_{\alpha, \sigma_1, \sigma_2}\, a^\dagger_{\alpha, \sigma_1}\, 
\boldsymbol{\sigma}^{\phantom \dagger}_{\sigma_1, \sigma_2}\, a^{\phantom \dagger}_{\alpha, \sigma_2}\ ,
\end{equation}  
and the charge operator reads 
\begin{equation}
N= \sum_{\alpha, \sigma}\, a^\dagger_{\alpha,\sigma} a^{\phantom \dagger}_{\alpha,\sigma} \ .
\end{equation}
In (\ref{eq:Hdot}) we neglect, {\it inter alia}, the magnetic anisotropy, which is frequently present in real systems.
The external magnetic field is also chosen along the $z$-axis $\mathbf{B}=(0,0,B)$. For the tunneling part of the Hamiltonian, we choose
\begin{equation}
H_{tun.} =\sum_{n=1}^{N_l + N_r}\sum_{\alpha,\sigma} \int \frac{dk}{2\pi}\, t^{\phantom\dagger}_{\alpha, n}\, a^\dagger_{\alpha, \sigma} c^{\phantom\dagger}_{nk, \sigma} + h.c.\ .
\end{equation} 
Note that we have chosen 
a diagonal form of the single particle part of $H_{dot}$. This means that the tunneling amplitudes $t_{\alpha,n}$ are random (at least the signs are). 

The system dynamics is fully defined by specifying the distribution functions of electrons in the leads. Note that the distribution function of the dot cannot be chosen freely. Due to the coupling to the leads the system will "forget" its initial distribution function in favor of that imposed by the leads on a time scale of the order of the relaxation time. The latter 
can be estimated as the inverse level-broadening of the dot states. The situation with applied voltage is described by different electrochemical potentials on the left lead and the right lead. The distribution functions of the leads are thus chosen as $n_l(\epsilon)=1/(e^{\beta (\epsilon - (\mu +V))}+1)$ and $n_r(\epsilon)=1/(e^{\beta (\epsilon - \mu)}+1)$, where $\mu$ is the bare chemical potential. Note that we assume the bare chemical potential to be the same for both leads, i.e., $V$ is the applied voltage.

\section{Derivation of the effective action}
\label{sec:Derivation}

\subsection{Decoupling of the interactions} 
\label{eq:Decoupling}

We employ the Keldysh formalism, since we consider a system driven out of equilibrium. We use the path integral technique. The Keldysh generating function (we do not specify the source fields) is then given by  
\begin{equation}
\mathcal{Z}=\int D[\bar \Psi, \Psi] e^{i \mathcal{S}[\bar \Psi, \Psi]}\ .
\end{equation} 
The Keldysh action reads 
\begin{equation}
i \mathcal{S} = i \oint_K dt\,  \left[\bar \Psi\, i \partial_t\, \Psi - H(\bar \Psi , \Psi) \right]\ ,
\end{equation} 
where $\bar \Psi, \Psi$ denote the fermionic fields, and the Keldysh integral 
runs from $-T_K$ to $+T_K$ on the upper part of the contour ($+$) and in reverse over the lower part of the contour ($-$). 
We use Hubbard-Stratonovich transformations to decouple the interaction. In particular, we substitute 
\begin{equation}
e^{i J \oint dt\, \mathbf{S}^2} = \int D\mathbf B_\mathrm{exc}\, e^{-i \oint dt\, \left(\frac{\mathbf B_\mathrm{exc}^2}{4J} - \mathbf B_\mathrm{exc} \mathbf{S}\right)}
\end{equation} 
for the exchange interaction, while  for the Coulomb interaction we take
\begin{equation}
e^{-i E_{c} \oint dt\, (N-N_0)^2} = \int D V_d\, e^{i \oint dt\, \left(\frac{V_d^2}{4E_{c}} - V_d (N-N_0)\right)}\ .
\end{equation} 
We note that 
$\mathbf B_\mathrm{exc}$ is the exchange field generated on the dot.
It is proportional to the magnetization, as can be seen by variation of the action with respect to the quantum component of $\mathbf B_\mathrm{exc}$, which 
leads to the condition $\mathbf B_\mathrm{exc}^c = 2J \langle \mathbf S \rangle$ (the superscript $c$ denotes the classical Keldysh component). Analogously, $V_d$ is the electrostatic potential generated by electrons on the dot.
Variation of the action with respect to the quantum component of $V_d$, i.e. $V_d^q$, gives $V_d^c= 2 E_{c} (\langle N \rangle -N_0)$. Following the Hubbard-Stratonovich transformation the action is quadratic in the fermionic fields. The latter can be integrated out, and the resulting determinant can be 
reexponentiated leading to 
\begin{eqnarray}
i \mathcal{S}_{\mathbf{M}, V_d} &=& \mathrm{tr}\, \mathrm{ln}\, \left[ -i \left( 
\begin{array}{ccc}
G_l^{-1} & -t_l^\dagger & 0 \\ 
-t_l & G_{d}^{-1} +\mathbf{M} \frac{\boldsymbol \sigma}{2}- V_d & -t_r \\ 
0 & -t_r^\dagger & G_r^{-1}
\end{array} 
\right) \right] \nonumber \\*
& & \hspace{-3.5em} - i \oint_K dt\, \frac{(\mathbf M - \mathbf B )^2}{4J} +i \oint_K dt\, \left(\frac{V_d^2}{4E_{c}} + V_d N_0 \right)\ ,
\end{eqnarray}
where we introduced $\mathbf M = \mathbf B + \mathbf B_\mathrm{exc}$ to which we refer to in the following as the magnetization~\footnote{$\mathbf B_\mathrm{exc}$ is proportional to the true magnetization and, in the ferromagnetic case, $|\mathbf B_\mathrm{exc}| \gg |\mathbf{B}|$ and thus $\mathbf{M} \approx \mathbf B_\mathrm{exc}$. This justifies to refer to $\mathbf M$ as the magnetization.}. The Green's functions are defined as $G_{d}^{-1}=  i \partial_t -H_{dot,0}$  for the dot and $G_{l,r}^{-1}= i \partial_t - H_{l,r}$ 
for the leads.
By expanding and resumming the $\mathrm{tr}\, \mathrm{ln}\, [ ... ]$  in the tunneling matrices we obtain 
\begin{eqnarray}
\hspace{-2em}i \mathcal{S}_{\mathbf{M}, V_d} & = & \mathrm{tr}\, \mathrm{ln}\, \left[ -i \left( G_{d}^{-1} +\mathbf{M} \frac{\boldsymbol \sigma}{2}- V_d - \Sigma \right) \right] \nonumber \\*
& & \hspace{-3.5em} - i \oint_K dt\, \frac{(\mathbf M- \mathbf B)^2}{4J} +i \oint_K dt\, \left(\frac{V_d^2}{4E_{c}} + V_d N_0 \right)\ .
\end{eqnarray}
Here we defined $\Sigma=\Sigma_l + \Sigma_r$, where $\Sigma_l = t_l^{\phantom\dagger} G_l^{\phantom\dagger} t_l^\dagger$ and $\Sigma_r = t_r^{\phantom\dagger} G_r^{\phantom\dagger} t_r^\dagger$. We have dropped the terms $\mathrm{tr}\, \mathrm{ln}\, \left[ -i G_l^{-1} \right]$ and $\mathrm{tr\, ln}\, \left[ -i G_r^{-1} \right]$, since they do not contain any source fields. The time dependence of $\mathbf{M}$ and $V_d$ renders the $\mathrm{tr}\, \mathrm{ln}[...]$ part of the action complicated. However, by applying a number of gauge transformations, we can transform the action to the rotating frame and, thus, shift the time dependence to the self-energies. This procedure leads to the AES type of effective action and allows us to make further progress.

\subsection{The rotating frame} 

We perform a rotation in spin space, such that $\mathbf{M}$ becomes parallel to the z-axis. We closely follow the ideas of Ref~\cite{PRLShnirman2015}. That is, we split $\mathbf{M}=M \mathbf n$ into an amplitude $M$ and a direction $\mathbf n$. Then we introduce spin-rotation matrices $R$, such that $\mathbf{n} \boldsymbol \sigma = R \sigma_z R^\dagger$. We use the Euler angle representation $R= e^{-i \frac{\phi}{2} \sigma_z} e^{-i \frac{\theta}{2} \sigma_y} e^{i \frac{\phi - \chi}{2} \sigma_z}$. The rationale behind this choice is that by construction $\theta_+(-T_K)=\theta_-(-T_K)$ 
between the upper ($+$) and lower ($-$) parts of the contour, and similarly $\phi_+(-T_K)=\phi_-(-T_K)+2\pi p$, where $p$ is an integer. Thus, the boundary condition $R_{+}(-T_K)=R_{-}(-T_K)$ is satisfied for any choice of the gauge field 
$\chi(t)$ such that $\chi_+(-T_K) = \chi_-(-T_K) + 4\pi m$. The integer $m$ can be chosen arbitrarily.

The time dependent transformation rotating $\mathbf M$ to point along the z-axis generates the usual geometric term $Q\equiv - i R^\dagger \dot R= Q_\parallel + Q_\perp$, where $Q_\parallel \equiv (1/2)[\dot \phi\, (1-\cos \theta ) - \dot \chi] \sigma_z$, and 
$
Q_\perp \equiv 
(1/2)\exp{\left[i\chi\sigma_z\right]}\,\left[\dot \phi \sin\theta \,\sigma_x -\dot \theta \sigma_y \right] \,\exp{\left[i\phi\sigma_z\right]}
$. 
Now, the action reads
\begin{eqnarray}
\hspace{-2em}i \mathcal{S}_{\mathbf{M}, V_d} & = & \mathrm{tr}\, \mathrm{ln}\, \left[ -i \left( G_{d,z}^{-1}- Q - V_d - R^\dag \Sigma R \right) \right] \nonumber \\*
& & \hspace{-3.5em} - i \oint_K dt\, \frac{(\mathbf M- \mathbf B)^2}{4J} +i \oint_K dt\, \left(\frac{V_d^2}{4E_{c}} + V_d N_0 \right)\ ,
\end{eqnarray}
where $G_{d,z}^{-1} = i \partial_t -H_{dot,0} + M \frac{\sigma_z}{2}$ is the rotated Green function of the dot.

\subsection{Ferromagnetic regime} 

In the following, we consider only itinerant ferromagnets well beyond the Stoner transition, although our analysis may apply also near the Stoner transition. That is, we assume that a large magnetization $M$ is built up on the dot. Furthermore, we assume adiabaticity, meaning that  
the angular velocity of the magnetization rotation is much smaller than the magnetization
(measured in units of $g |e|/(2mc)$). Thus, in $Q$ we only keep the Berry phase part $Q_\parallel$, whereas we neglect the part $Q_\perp$ responsible for single-electron Landau-Zener transitions. In the deep ferromagnetic phase 
fluctuations of the length of the magnetization, $\delta M$, are small compared to $M$ (cf. Ref.~\cite{PRBBaskoVavilov2009}). In the following, we will assume that the magnetization has a fixed length, $M_0$.

\subsection{Choice of gauge}
Ideally, we would have liked to choose $\dot \chi = \dot \phi (1- \cos \theta)$ as a gauge on the entire contour, since this would eliminate $Q_\parallel$~\cite{PRLShnirman2015} (from now on we identify $Q$ with $Q_\parallel$ as 
$Q_\perp$ is neglected). However, we have to respect the boundary conditions $\chi_+(-T_K)-\chi_-(-T_K)=4  \pi m$. Due to these boundary conditions we cannot eliminate the quantum component $Q^q$. We can eliminate though the classical component by choosing a gauge such that $Q_+=-Q_-$. This is achieved by 
\begin{equation}
\dot \chi_\pm = {\dot \phi}^c p^c + \frac{1}{4} \dot \phi^q p^q \pm \left( \frac{1}{2} {\dot \phi}^q p^c + \frac{1}{2} \phi^q 
{\dot p}^c \right)\ ,
\end{equation}
where $p=(1-\cos \theta )$ and we have chosen the specific boundary condition $m=0$. For the quantum part we obtain 
$Q^q= Q_+-Q_-= (1/2)\left( {\dot \phi}^c p^q - \phi^q {\dot p}^c\right) \sigma_z$ . Up to second order in the quantum components of $\phi$ and $\theta$, this choice is the same as in Ref.~\cite{PRLShnirman2015}. 

\subsection{U(1) gauge transformation} 

Besides rotating the magnetization, we also transfer the potential $V_d$ to the self energy part by a U(1) gauge transformation $e^{-i \psi}$ (this is the original AES transformation~\cite{AES_PRL,AES_PRB}). We would like to eliminate the complete potential by the choice $\dot \psi= V_d$ on the Keldysh contour. However, we have to respect the boundary conditions again, i.e. $\psi_-(-T_K)- \psi_+(-T_K)= 2 \pi k$, with integer $k$. And again, in this semiclassical limit, in which charge quantization is neglected (strong tunnel coupling between the dot and one or both leads), we can restrict ourselves to a single value of $k$, e.g., $k=0$. We, therefore, do not gauge out the zero frequency quantum component $V_{d\, 0}^q$ (cf. Ref.~\cite{KamenevGefenZeroMode}), since $\psi_-(-T_K)- \psi_+(-T_K)= \oint_K dt\, V_d = \int_{-T_K}^{T_K} dt\, (V_d^+-V_d^-) = \int_{-T_K}^{T_K} dt\, V_d^q= V_{d\, 0}^q$. Our choice for the gauge field is
\begin{equation}
\begin{array}{ccc}
\dot \psi_+ & = & V_d^+ -\frac{1}{2} V_{d\, 0}^q \\ 
\dot \psi_- & = & V_d^- +\frac{1}{2} V_{d\, 0}^q 
\end{array} \label{U(1) gauge}
\end{equation}
which leaves the zero frequency quantum component $V_{d\, 0}^q$ untouched but gauges out the rest of $V_d$.

\subsection{Keldysh rotation}

Employing the combined $U(1)\times SU(2)$ gauge transformations with $U=R\, e^{-i \psi}$ and introducing the Keldysh 
matrix structure we obtain the following effective action
\begin{eqnarray}
i \mathcal{S}_{\mathbf{M}, V_d}  &=&  \mathrm{tr}\, \mathrm{ln}\, \left[ -i \left( \tau_z G_{d,z}^{-1}-   \frac{Q^q}{2} \tau_0 - \frac{V_{d\, 0}^q }{2} \tau_0 - \hat U^\dagger \hat \Sigma \hat U \right) \right]\nonumber \\ &+& i \int dt\, \frac{\mathbf{B} \mathbf{M}^q}{2J} + i \int dt\, \left( \frac{V_d^c V_d^q}{2 E_{c}} + V_d^q N_0\right)\ , \nonumber\\ \label{action Keldysh not rotated}
\end{eqnarray}
where $\hat\Sigma =  \left(\begin{array}{cc}
\Sigma_{++} & -\Sigma_{+-} \\ 
-\Sigma_{-+} & \Sigma_{--}
\end{array}\right)$ and $\hat U = 
\left(\begin{array}{cc}
U_+ & 0 \\ 
0 & U_-
\end{array}\right)
=
U^c\, \tau_0 + (1/2) U^q\, \tau_z$.
Finally, we perform the standard "bosonic" rotation~\cite{KamenevBook}
with 
$
L= \frac{1}{\sqrt{2}} \left(\begin{array}{cc}
1 & 1 \\ 
1 & -1
\end{array}\right)
$ and obtain
\begin{eqnarray}
i \mathcal{S}_{\mathbf{M}, V_d}  &=&  \mathrm{tr}\, \mathrm{ln}\, \left[ -i \left( \tilde G_{d,z}^{-1}-   \frac{Q^q}{2} \tau_0 - \frac{V_{d\, 0}^q }{2} \tau_0  - \tilde U^\dagger \tilde \Sigma \tilde U \right) \right]  \nonumber\\ &+&  i \int dt\, \frac{\mathbf{B} \mathbf{M}^q}{2J} + i \int dt\, \left( \frac{V_d^c V_d^q}{2 E_{c}} + V_d^q N_0\right)\ ,\nonumber\\ \label{action}
\end{eqnarray}
where  $\tilde U \equiv L^\dag\hat U L = U^c\, \tau_0 + (1/2) U^q\, \tau_x$, 
$\tilde G_{d,z}^{-1} \equiv L^\dag \tau_z G_{d,z}^{-1} L=  \tau_x G_{d,z}^{-1}$, and
$\tilde \Sigma \equiv L^\dag \hat\Sigma L= \left(
 \begin{array}{cc}
 0 & \Sigma^A \\ 
 \Sigma^R & \Sigma^K
 \end{array} \right)$.

\section{Stationary quasi-classical trajectories}
\label{sec:Solution}

\subsection{General considerations}

A saddle point trajectory of magnetization and voltage $U_{sp}(t)$ can be found by considering a variation,
$U(t) = U_{sp}(t) + \delta U(t)$, expanding the effective action in components of $\delta U$ and requiring the linear order 
of the expansion to vanish. This strategy, in the laboratory frame, was pursued  in Ref.~\cite{PRBBaskoVavilov2009}. 
For semiclassical trajectories $U_{sp}(t)$ would be purely classical (no quantum components) and the expansion (up to the linear order) is effectively in the quantum components of $U(t)$. (More general saddle point solutions, e.g., instantons~\cite{PRBTitovGutman2016}, would require special care.) The expansion can be performed by splitting the Keldysh rotated (see Eq.~(\ref{action})) self-energy $\tilde U^\dagger \tilde \Sigma \tilde U$ into the saddle point part 
$U^\dagger_{sp} \tilde \Sigma U_{sp}$ and the rest, i.e.
\begin{widetext}
\begin{eqnarray}
i \mathcal{S}_{\mathbf{M}, V_d} & = &  \mathrm{tr}\, \mathrm{ln}\, \left[ -i \left( \tilde G_{d,z}^{-1}- U^\dagger_{sp} \tilde \Sigma U_{sp} -   \frac{Q^q}{2} \tau_0 - \frac{V_{d\, 0}^q }{2} \tau_0 - \left( \tilde U^\dagger \tilde \Sigma \tilde U -U^\dagger_{sp} \tilde \Sigma U_{sp} \right) \right) \right] \nonumber \\*
& + & i \int dt\, \frac{\mathbf{B} \mathbf{M}^q}{2J} + i \int dt\, \left( \frac{V_d^c V_d^q}{2 E_{c}} + V_d^q N_0\right)\ . \label{action cl-cl separated}
\end{eqnarray}
\end{widetext}
To perform the expansion in $Q^q$, $V_{d\, 0}^q$, and $ \tilde U^\dagger \tilde \Sigma \tilde U -U^\dagger_{sp} \tilde \Sigma U^{\phantom \dag}_{sp}$ one would have to find the zeroth order Green's function 
\begin{equation}\label{eq:full kinetic}
\tilde G_d \equiv (\tilde G_{d,z}^{-1}- U^\dagger_{sp} \tilde \Sigma U_{sp})^{-1}\ .
\end{equation} 
For a general non-stationary trajectory $U_{sp}(t)$ this task is akin to solving a time-dependent kinetic equation.
Then, expanding, we obtain the following effective action 
\begin{eqnarray}
i \mathcal{S}_{\mathbf{M}, V_d} & = &  i \mathcal{S}_{WZNW} +  i \mathcal{S}_{zero\ mode}  + i \mathcal{S}_{AES}
\nonumber \\* 
& & \hspace{-3.5em} + i \int dt\, \frac{\mathbf{B} \mathbf{M}^q}{2J} + i \int dt\, \left( \frac{V_d^c V_d^q}{2 E_{c}} + V_d^q N_0\right)\ . \label{action expanded}
\end{eqnarray}
Here the first term is the standard Berry phase action also known as Wess-Zumino-Novikov-Witten (WZNW) action, 
\begin{equation}\label{eq:WZNWgeneral}
i \mathcal{S}_{WZNW}= - \frac{1}{2} \int dt\, \mathrm{tr} [ \tilde G_d^K(t,t) Q^q(t)] \ .
\end{equation}
The second term is given by 
\begin{equation}
i \mathcal{S}_{zero\ mode}=  - \frac{1}{2} \int dt\, \mathrm{tr} [ \tilde G_d^K(t,t) V_{d\,0}^q] \ .
\end{equation}
Finally, the third term of Eq.~(\ref{action expanded}) is the AES action:
\begin{eqnarray}
\label{eq:AESgeneral}
i \mathcal{S}_{AES} & = & - \mathrm{tr} \left[ \tilde G_d \left(\tilde U^\dagger \tilde \Sigma \tilde U -  U_{sp}^\dagger \tilde \Sigma  U_{sp}  \right) \right] \nonumber \\*
&=& - \mathrm{tr} \left[ \tilde G_d \tilde U^\dagger \tilde \Sigma \tilde U \right]\ .
\end{eqnarray}
Curiously, the term $U_{sp}^\dagger \tilde \Sigma U_{sp}$ drops due to $U_{sp}$ being purely classical.

\subsection{Non-equilibrium stationary distribution function} 

Solving the kinetic equation (\ref{eq:full kinetic}) for an arbitrary (yet unknown) saddle point trajectory, $U_{sp}$, may be rather complicated (in Ref.~\cite{PRBBaskoVavilov2009} this has been achieved assuming very slow dynamics of the magnetization). We thus pursue a less ambitious task: we presume a stationary solution, solve the stationary kinetic equation corresponding to this solution  and check the presumed 
solution for stability and consistency. By doing so we consider only the long time limit so that the electron distribution 
function of the dot has already adjusted itself.   

We assume the stationary solution of the form $\theta(t) = \theta_0$, $\dot \phi(t)=-B_0$, and $\dot\psi(t)= V_d(t) = V_{d0}$ (if $\sin\theta_0=0$ the choice of $\dot\phi$ is immaterial). These relations hold on both parts of the Keldysh contours (the upper and the lower), we are thus talking about a purely classical trajectory. 
We introduce a purely classical $U_0$, which emerges from $U$ under the substitutions $\theta(t) \rightarrow \theta_0$, $\phi(t) \rightarrow -B_0\, t$, and $\psi (t) \rightarrow V_{d0}\, t$. 
Next we rewrite Eq.~(\ref{eq:full kinetic}) as
\begin{equation}\label{eq:full kinetic expanded}
\tilde G_d \equiv \left(\tilde G_{d,z}^{-1} - U_0^\dagger \tilde \Sigma U_0 - (U^\dagger_{sp} \tilde \Sigma U_{sp}-U_0^\dagger \tilde \Sigma U_0)\right)^{-1}\ .
\end{equation} 
Formally, we can expand in 
$(U^\dagger_{sp} \tilde \Sigma U_{sp}-U_0^\dagger \tilde \Sigma U_0)$, i.e.,
\begin{equation}\label{eq:kinetic expanded}
\tilde G_d = \tilde G_{d0} + 
\tilde G_{d0}(U^\dagger_{sp} \tilde \Sigma U_{sp}-U_0^\dagger \tilde \Sigma U_0)\tilde G_{d0} + \dots\ ,
\end{equation}
where $\tilde G_{d0} \equiv  \left(\tilde G_{d,z}^{-1} - U_0^\dagger \tilde \Sigma U_0 \right)^{-1}$.
We keep, however, only the lowest order term and approximate
\begin{equation}\label{eq:kinetic0}
\tilde G_d \approx \tilde G_{d0}\ .
\end{equation} 
This approximation is definitely sufficient for finding the stationary trajectories, since in this case 
$U_{sp} = U_0$. The validity of this approximation for the stability analysis of the stationary solutions will be discussed below. 

We now solve Eq.~(\ref{eq:kinetic0}). We only keep the contributions diagonal in spin and orbital space. To neglect the spin off-diagonal contributions is justified, since we assume $M_0$ to be the largest energy scale in the problem. The randomness in the coupling to the leads as well as the assumption of the large number of weakly conducting (tunneling) transverse channels justify the dropping of the orbital off-diagonal contributions (see Appendix \ref{sec:AppOffDiag}). The system is, thus, described by just three tunneling rates $\Gamma_l^\uparrow, \Gamma_l^\downarrow, \Gamma_r$ (see Appendix \ref{sec:AppOffDiag}). Here $\Gamma_l^{\sigma}$ is the tunneling rate (inverse life time) of an orbital state 
to states of the left lead with spin projection $\sigma$. Analogously $\Gamma_r = \Gamma_r^\uparrow = \Gamma_r^\downarrow$ is the (spin resolved) inverse life time due to the right lead.  

\begin{figure*}
\begin{center}
\includegraphics[width=0.5\textwidth]{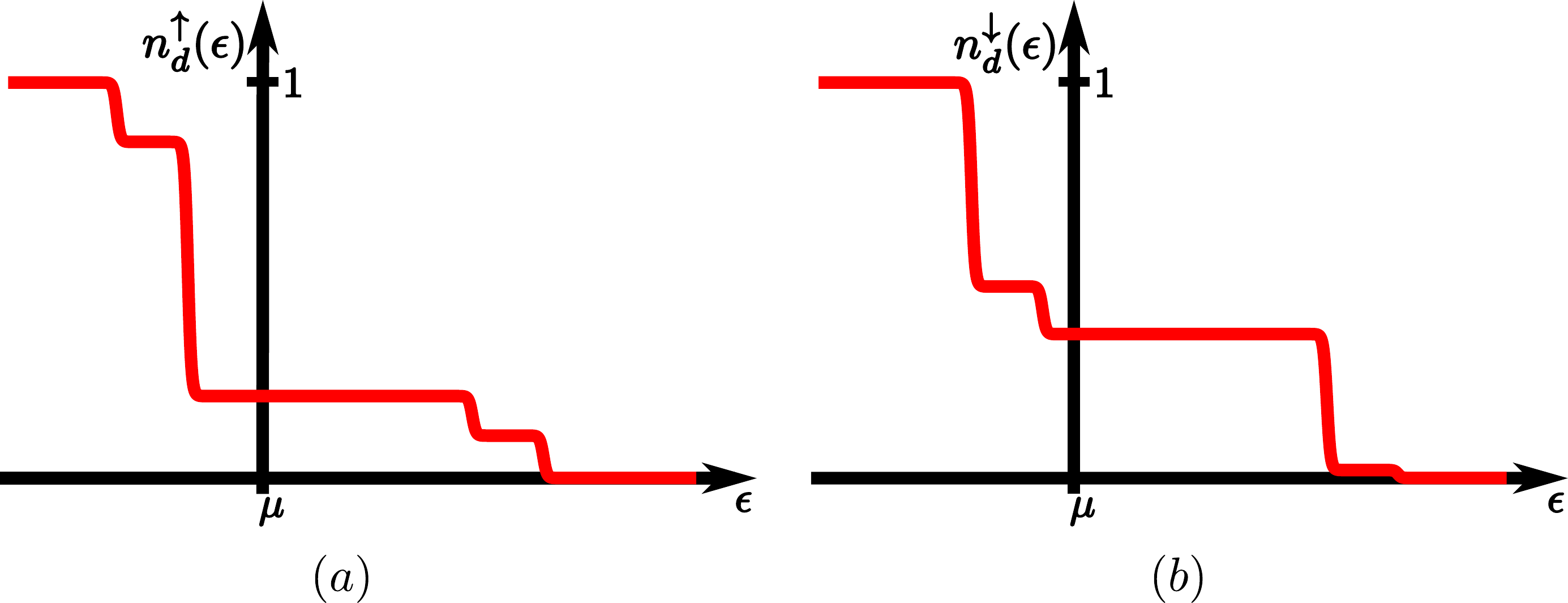}
\end{center}
\caption{(color online) The figure presents schematically the distribution function of Eq.~(\ref{Fd}). Here
$n_d^\sigma(\epsilon) \equiv (1-F_d^\sigma(\epsilon))/2$.  \label{fig:Fd}}
\end{figure*}

We obtain for the Green's function
\begin{equation}
\tilde G_d = \left( \begin{array}{cc}
G_d^K & G_d^{R} \\ 
G_d^A & 0
\end{array} \right)\ ,
\end{equation}
where 
\begin{equation}
G_d^{R/A}=\frac{1}{\epsilon - \epsilon_\alpha - (M_0/2) \sigma \pm i \Gamma_\sigma(\theta_0)}\ .
\label{GdRA}
\end{equation} 
Here $\Gamma_\sigma(\theta_0) = \cos^2 \frac{\theta_0}{2} \Gamma_l^\sigma + \sin^2 \frac{\theta_0}{2} \Gamma_l^{\bar \sigma} + \Gamma_r $ and $\bar\sigma$ denotes the spin projection opposite to $\sigma$.  Using the Ansatz $G_d^K= G_d^R F_d - F_d G_d^A$, we obtain the following spin-dependent distribution function:
\begin{eqnarray}
\label{Fd}
F_d^\sigma(\epsilon) = \frac{1}{\Gamma_\sigma (\theta_0)} & \Bigg[ & \cos^2 \frac{\theta_0}{2} \Gamma_l^\sigma F(\epsilon - \sigma B_-+V_{d0} - V) \nonumber \\
    & + & \sin^2 \frac{\theta_0}{2} \Gamma_l^{\bar \sigma} F(\epsilon - \bar \sigma B_+ +V_{d0} - V) \nonumber \\
    & + & \cos^2 \frac{\theta_0}{2} \Gamma_r F(\epsilon - \sigma B_- +V_{d0}) \nonumber \\
    & + & \sin^2 \frac{\theta_0}{2} \Gamma_r F(\epsilon - \bar \sigma B_+ +V_{d0}) \Bigg]
\ .
\end{eqnarray}
Here we introduce $F(\epsilon)\equiv \tanh \left( \frac{\epsilon - \mu}{2 T}\right)$ and $B_{\pm} \equiv B_0 (1\pm \cos\theta_0)/2$.
The strong non-equilibrium character of the distribution function in the rotating frame (which represents
far from equilibrium conditions in the laboratory frame) is obvious: it is a sum of four equilibrium distribution functions with different energy shifts. We demonstrate this schematically in Fig.~\ref{fig:Fd}.
Note that this "distribution function" is an auxiliary quantity defined in the rotating frame, and is not necessarily equal to the actual distribution function (Keldysh Green's function) of the dot.

\subsection{The WZNW and the AES terms of the action}

For the Berry-phase (WZNW) action (\ref{eq:WZNWgeneral}) we obtain
\begin{eqnarray}\label{eq:WZNW}
i \mathcal{S}_{WZNW} & = & - \frac{1}{2} \int dt\, \mathrm{tr} [ \tilde G_d^K(t,t) Q^q(t)] \nonumber \\
& = & -i S \int dt\, \sin \theta^c \left( \theta^q \dot \phi^c - \phi^q \dot \theta^c \right) \ ,
\end{eqnarray}
where we introduced the total spin length $S$, which is about half the number of orbital states between $\mu - M_0/2$ and $\mu + M_0/2$ (we assume low enough temperature, $k_{\rm B} T  \ll M_0$). For details of the calculation, see Appendix \ref{sec:calc.WZNW}. The self consistency discussed in Sec.~\ref{eq:Decoupling} requires $M_0 \approx 2JS$.

Now we evaluate the AES-like action given by Eq.~(\ref{eq:AESgeneral}).
One significant difference between the AES-like action $i \mathcal{S}_{AES}$ and the original AES action~\cite{AES_PRL,AES_PRB} is the spin-structure under the trace. To handle this complication, we introduce
\begin{eqnarray}
U & = & A_{\uparrow \uparrow} \sigma_\uparrow + A_{\downarrow \downarrow} \sigma_\downarrow + A_{\downarrow \uparrow} \sigma_+ + A_{\uparrow \downarrow} \sigma_-\ , \nonumber \\ 
U^\dagger & = & A_{\uparrow \uparrow}^* \sigma_\uparrow + A_{\downarrow \downarrow}^* \sigma_\downarrow + A_{\downarrow \uparrow}^* \sigma_- + A_{\uparrow \downarrow}^* \sigma_+\ ,
\end{eqnarray}
where $\sigma_{\uparrow / \downarrow}= \frac{1}{2} (\sigma_0 \pm \sigma_z)$, $\sigma_\pm = \frac{1}{2} (\sigma_x 
\pm i\sigma_y)$ and 
\begin{eqnarray}
A_{\uparrow \uparrow} & = & \cos\frac{\theta}{2}\, e^{ i ( -\frac{\chi}{2} - \psi)}\ , \qquad A_{\downarrow \uparrow}  =  - \sin \frac{\theta}{2}\,  e^{i (-\phi + \frac{\chi}{2} -\psi)}\ , \nonumber \\*
A_{\downarrow \downarrow} & = & \cos \frac{\theta}{2} \, e^{i ( \frac{\chi}{2} -\psi)}\ , \qquad A_{\uparrow \downarrow} =\sin \frac{\theta}{2}\, e^{i ( \phi - \frac{\chi}{2} -\psi)} \ .
\end{eqnarray}
Now, we can explicitly take the trace over the spin-space to obtain
\begin{equation}
i \mathcal{S}_{AES}= - \sum_{\sigma \sigma'} \mathrm{tr} \left[ \tilde G_d(\sigma)\,  \tilde A^*_{\sigma \sigma'} \tilde \Sigma_{\sigma'} \tilde A_{\sigma \sigma'} \right]\ ,
\end{equation}
which resembles 4 copies of the AES-problem, one for each combination of $\sigma, \sigma'$. Note, however, that the four combinations $A_{\sigma \sigma'}$ are not independent, since they only describe three coordinates $\phi , \psi, \theta$.  The calculation of the AES-like action is straightforward (for details see Appendix \ref{sec:calc.AES}) and 
we obtain $i \mathcal{S}_{AES} =  i \mathcal{S}_{AES}^R +  i \mathcal{S}_{AES}^K$, where the 
retarded part and the Keldysh part are given by 
\begin{eqnarray}\label{eq:SAESR}
i \mathcal{S}_{AES}^R & = & -i \int dt\, dt'\, \nonumber \\*
& \times & \sum_{\sigma \sigma'} \mathrm{Im} \left[ A^*_{c, \sigma \sigma'}(t')\, \alpha^{R}_{\sigma \sigma'}(t-t')\,  A_{q, \sigma \sigma'} (t)\right]\ , \nonumber\\  
\end{eqnarray}
\begin{eqnarray}\label{eq:SAESK}
i \mathcal{S}_{AES}^K & = & - \frac{1}{4} \int dt\, dt'\, \nonumber \\* 
& \times & \sum_{\sigma \sigma'} A^*_{q, \sigma \sigma'}(t')\, \alpha^{K}_{\sigma \sigma'}(t-t')\,  A_{q, \sigma \sigma'} (t)\ .
\end{eqnarray}
The retarded kernel, $\alpha^{R}_{\sigma \sigma'}(\epsilon)= g^{\sigma \sigma'}_l \alpha^R_{l,\sigma}(\epsilon)+ g^{\sigma \sigma'}_r \alpha^R_{r,\sigma}(\epsilon)$, and the Keldysh kernel, $\alpha^{K}_{\sigma \sigma'}(\epsilon)= g^{\sigma \sigma'}_l \alpha^K_{l,\sigma}(\epsilon)+ g^{\sigma \sigma'}_r \alpha^K_{r,\sigma}(\epsilon)$, have contributions from both leads.
The conductances are given by $g^{\sigma \sigma'}_{l} = 2 \rho_d^\sigma \Gamma^{\sigma'}_{l}$ and 
$g^{\sigma \sigma'}_{r} = 2 \rho_d^\sigma \Gamma_{r}$ ($g^{\sigma \sigma'}_{r}$ is independent of $\sigma'$), where $\rho_d^\sigma$ is the spin dependent density of states of the dot (see Appendix~\ref{sec:calc.AES}).
The retarded kernel functions are given by 
\begin{eqnarray}
\label{alphal}
\mathrm{Re}\left[\alpha_{l,\sigma}^R(\epsilon)\right] & = &
\frac{1}{2}\,\int d\omega \left[F^{\sigma}_d(\omega+\epsilon)-F_l(\omega)\right] \nonumber \\* 
& = & \epsilon + V + \tilde V_\sigma (\theta_0)\ ,
\end{eqnarray}
\begin{eqnarray}
\label{alphar}
\mathrm{Re}\left[\alpha_{r,\sigma}^R(\epsilon)\right] & = & \frac{1}{2}\,\int d\omega \left[F^{\sigma}_d(\omega+\epsilon)-F_r(\omega)\right] \nonumber \\* & = & \epsilon + \tilde V_\sigma (\theta_0)\ ,
\end{eqnarray}
where 
\begin{equation}\label{Vsigma}
\tilde V_\sigma (\theta_0) = V_{d0} - V +  \frac{\Gamma_r V - \Gamma_\Delta \sin^2\theta_0 \, \frac{B_0}{2}}{\Gamma_\sigma(\theta_0)}\ . 
\end{equation} 
Here $\Gamma_\Delta \equiv \frac{1}{2}(\sigma \Gamma_l^\sigma + \bar \sigma \Gamma_l^{\bar\sigma}) =\frac{1}{2}( \Gamma_l^\uparrow - \Gamma_l^{\downarrow})$. 
We disregard the imaginary parts of the kernel functions, since these contributions lead to a renormalization of the magnetic field, which is included in $B$. The Keldysh kernel functions are presented in  Appendix~\ref{sec:calc.AES}.

Note that had we enforced a Fermi distribution function, $F_d(\epsilon)=F(\epsilon)$, in the dot (in the rotating frame), 
which is also the distribution function of the right lead, 
we would have 
$\tilde V_\sigma (\theta_0)=0$ and the kernel functions (\ref{alphal},\ref{alphar}) would assume the usual AES form. 

In what follows we will need the following conductances (cf. Ref.~\cite{ChudnovskiyPRL}). The usual conductances for the charge current are given by $g_l(\theta)= \frac{\cos^2 (\frac{\theta}{2})}{4} (g^{\uparrow \uparrow}_l +g^{\downarrow \downarrow}_l) + \frac{ \sin^2 (\frac{\theta}{2})}{4} (g^{\uparrow \downarrow}_l +g^{\downarrow \uparrow}_l)$ for left contact and analogously for the right contact $g_r$. The conductances related to the dissipation of the magnetization read $\tilde g_l(\theta)= \frac{\sin^2 (\frac{\theta}{2})}{4} (g^{\uparrow \uparrow}_l +g^{\downarrow \downarrow}_l) + \frac{ \cos^2 (\frac{\theta}{2})}{4} (g^{\uparrow \downarrow}_l +g^{\downarrow \uparrow}_l)$ for left contact and analogously for the right contact $g_r$. The spin-conductance is defined as $g_s= \frac{1}{4} (g^{\uparrow \uparrow}_l - g^{\downarrow \downarrow}_l - g^{\uparrow \downarrow}_l + g^{\downarrow \uparrow}_l)$. Note that there is no spin conductance appearing for the right lead, since it is not magnetic. It also follows that $g_r$ is independent of $\theta$ and furthermore $\tilde g_r = g_r$.

\subsection{Quasi-classical equations of motion} 

The quasi-classical equations of motion can be determined by variation 
with respect to the quantum components~\footnote{The resulting equations of motion are for the classical components of 
the corresponding fields. We drop the superscript '$c$' for brevity.}, $\theta^q, \phi^q, \psi^q$. In this paper we
disregard the Keldysh part of the action (Eq.~(\ref{eq:SAESK})), which would render the equations of motion stochastic by generating Langevin terms~\cite{ASchmid82_Langevin,KamenevBook}.
For our setup with $\mathbf{B}= (0,0, B)$ and with the fixed magnetization of the left lead parallel to this field, and employing the self consistency relation $M_0 \approx 2J S$, we obtain the Landau-Lifshitz-Gilbert equation~\cite{Gilbert2004} supplemented by a spin-torque term
\begin{equation}
\frac{d \mathbf{M}}{d t}= -  \mathbf{B} \times \mathbf{M} - \alpha(\theta) \frac{\mathbf{M}}{M_0}\times \frac{d \mathbf{M}}{d t} + \frac{1}{S}\,\frac{\mathbf{M}}{M_0} \times \left( \mathbf{I}_s \times \mathbf{M} \right)\ . \label{LLG}
\end{equation}
Here the Gilbert dissipation coefficient $\alpha(\theta)= \frac{1}{S} (\tilde g_l(\theta) + \tilde g_r)$ depends on 
the angle $\theta$. Furthermore, we obtain Kirchhoff's law
\begin{equation}
C \dot V_d = I_l - I_r\ , \label{Kirchhoff}
\end{equation}
where $I_l$ is the current flowing into the dot from the left lead and $I_r$ is the current flowing out of the 
dot into the right lead (see Fig.~\ref{System}). The above equations of motion can be written explicitly in terms of the Euler angles.
They read
\begin{eqnarray}
\sin\theta\,\dot \phi & = & - \sin \theta B - \alpha(\theta)\, \dot \theta\ , \nonumber \\ 
\sin\theta\,\dot \theta & = &  \sin^2 \theta\, \left[ \alpha(\theta)\dot \phi - I_s/S \right]\ , \nonumber \\ 
C \dot V_d &=& I_l - I_r \ . \label{eom}
\end{eqnarray}

For the spin-torque current, we obtain
\begin{eqnarray}
I_s & = & g_s (V-V_d) + g_s (V_{d0}-V) \nonumber \\*
& + & S\,\alpha(\theta_0) \,  \frac{2 \Gamma_\Delta \Gamma_r V - \Gamma^2_\Delta \sin^2\theta_0 \, B_0 }{\Gamma_\Sigma^2 - \cos^2\theta_0 \Gamma_\Delta^2} \ ,  \label{SpinCurrent}
\end{eqnarray}
where $\Gamma_\Sigma \equiv (\Gamma_l^\uparrow + \Gamma_l^\downarrow)/2 + \Gamma_r$.
The last two terms in $I_s$ arise due to the non-equilibrium character of the electron distribution function $F_d^\sigma$ 
in the quantum dot/nano-particle (see Eq.~(\ref{Fd})). Had we enforced an equilibrium distribution by setting $F_d = F(\epsilon)$ in Eqs.~(\ref{alphal},\ref{alphar}), the spin-torque current would reduce to $I_s^{eq.}=g_s (V - V_{d})$
as, e.g., in Ref.~\cite{ChudnovskiyPRL}.

For the charge currents we obtain
\begin{eqnarray}
I_l &=&4g_l (V-V_d)- g_s \sin^2 \theta \dot \phi + 4g_l (V_{d0}-V) \nonumber \\* & + & \left[
\frac{\cos^2 \frac{\theta}{2}\,g_l^{\uparrow \uparrow} + \sin^2 \frac{\theta}{2}\,g_l^{\uparrow \downarrow}}{\Gamma_\uparrow(\theta_0)} +  \frac{\cos^2 \frac{\theta}{2}\, g_l^{\downarrow \downarrow}+ \sin^2 \frac{\theta}{2}\, g_l^{\downarrow \uparrow}}{\Gamma_\downarrow(\theta_0)}  \right]\nonumber \\* &  & \times  \left(\Gamma_r V - \Gamma_\Delta \sin^2\theta_0 \, \frac{B_0}{2}\right)\ , \label{LeftCurrent}
\end{eqnarray}
and
\begin{eqnarray}
I_r &=&4g_r V_{d} + 4 g_r (V -V_{d0}) \nonumber \\* &-& \left[
\frac{\cos^2 \frac{\theta}{2}\,g_r^{\uparrow \uparrow} + \sin^2 \frac{\theta}{2}\,g_r^{\uparrow \downarrow}}{\Gamma_\uparrow(\theta_0)} +  \frac{\cos^2 \frac{\theta}{2}\, g_r^{\downarrow \downarrow}+ \sin^2 \frac{\theta}{2}\, g_r^{\downarrow \uparrow}}{\Gamma_\downarrow(\theta_0)}  \right] \nonumber \\* & & \times \left(\Gamma_r V - \Gamma_\Delta \sin^2\theta_0 \, \frac{B_0}{2}\right)\ .  \label{RightCurrent}
\end{eqnarray}
Here again, the last two terms in $I_l$ and the last two terms in $I_r$ arise due to the non-equilibrium character of the distribution function and would therefore vanish for the enforced equilibrium distribution, i.e., $I_r^{eq.} = 4g_r V_{d}$ and $I_l^{eq.} = 4g_l (V-V_d)- g_s \sin^2 \theta \dot \phi$. For the left junction (magnetic tunnel junction), besides the ohmic term, the charge current $I_l^{eq.}$ contains also the current pumped by the precessing magnetization, $- g_s \sin^2 \theta \dot \phi$, as predicted 
in Refs.~\cite{BergerInverse,PRBTserkovnyak2008}. 

We emphasize that the structure of the Landau-Lifshitz-Gilbert equation (\ref{LLG}) and of the Kirchhoff's law (\ref{Kirchhoff}) 
is preserved, whereas the three currents, $I_s$, $I_l$, and $I_r$ are strongly modified due to the non-equilibrium character of the 
electron distribution function (\ref{Fd}). Furthermore, as the magnetization and the electrostatic potential of the dot are treated on an equal footing, we are able to describe their coupled dynamics. The coupling is mediated by the spin-torque current, $I_s$, and by the pumped 
current, $- g_s \sin^2 \theta \dot \phi$. The intimate relation between the two was always clear, but is now explicitly demonstrated as both are derived from the same effective action (see also~\cite{VirtanenHeikkila2016}).

\subsection{Stationary solutions}

The possible stationary solutions, $\theta_0,B_0,V_{d0}$, are obtained by substituting $\dot \phi =-B_0$, $V_d = V_{d0}$, $\theta = \theta_0$ into Eqs.~(\ref{eom}) and solving for $\theta_0,B_0,V_{d0}$. 
Assuming the stationary solution is such that $\sin\theta_0 \neq 0$, the first equation of Eqs.~(\ref{eom}) 
would immediately lead to $B_0 = B$. The second equation of (\ref{eom}) gives then 
\begin{equation}\label{eq:EqFortheta0}
0 = \alpha(\theta_0) \, \frac{\left[(\Gamma_\Sigma^2-\Gamma_\Delta^2)B + 2 \Gamma_r \Gamma_\Delta V\right]}{\Gamma_\Sigma^2 - \cos^2\theta_0 \Gamma_\Delta^2} \ .
\end{equation}
Since $\alpha(\theta_0)$ and $\Gamma_\Sigma^2 - \cos^2\theta_0 \Gamma_\Delta^2$ are always positive, 
there are no solutions with $\sin\theta_0 \neq 0$.
Thus only north and south pole solutions are possible. This is in contrast to the persistent precession ($\sin\theta_0\neq 0$) solutions discussed, e.g., in Ref.~\cite{ChudnovskiyPRL}. 

With $\theta = \theta_0$ and $\sin\theta_0 = 0$ the first equation of Eqs.~(\ref{eom}) is automatically 
satisfied for arbitrary $B_0$. This is because the dynamics of $\phi$ is meaningless for $\sin\theta_0 = 0$.
It is also easy to see, that the currents given by Eqs. \eqref{SpinCurrent}, \eqref{LeftCurrent}, \eqref{RightCurrent} become independent of $V_{d0}$ and, in turn, the third equation of Eqs.~(\ref{eom}) (Kirchhof's law) is satisfied for arbitrary value of $V_{d0}$. This fact will be discussed below (Sec.~\ref{subsec:RoleZeroMode}).
For the charge current, $I=I_l = I_r$, in the two stationary states we obtain
\begin{equation}\label{eq:I00}
I_{\theta_0=0} =\left[\frac{g_l^{\uparrow\uparrow} g_r^{\uparrow\uparrow}}
{g_l^{\uparrow\uparrow} + g_r^{\uparrow\uparrow}} + \frac{g_l^{\downarrow\downarrow} g_r^{\downarrow\downarrow}}
{g_l^{\downarrow\downarrow} + g_r^{\downarrow\downarrow}}\right]V \ ,
\end{equation}
and
\begin{equation}\label{eq:I0pi}
I_{\theta_0=\pi} =\left[\frac{g_l^{\uparrow\downarrow} g_r^{\uparrow\downarrow}}
{g_l^{\uparrow\downarrow} + g_r^{\uparrow\downarrow}} + \frac{g_l^{\downarrow\uparrow} g_r^{\downarrow\uparrow}}
{g_l^{\downarrow\uparrow} + g_r^{\downarrow\uparrow}}\right]V \ .
\end{equation}

\subsection{Stability of the stationary solutions}

\begin{figure*}
\begin{center}
\includegraphics[width=0.7\textwidth]{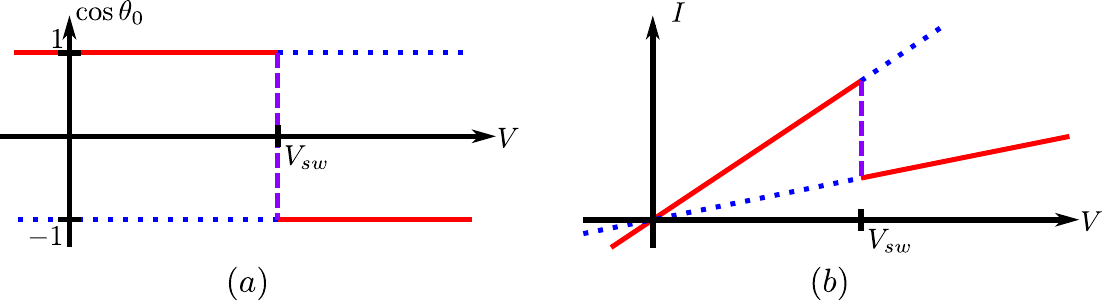}
\end{center}
\caption{(color online) The stationary solutions and their stability: solid red $\leftrightarrow$ stable, dotted blue $\leftrightarrow$ unstable, dashed purple $\leftrightarrow$ critical solutions. $(a)$ Here $\cos \theta_0$ is shown for applied $V$. For $V \neq V_{sw}$ the only stationary solutions are at the poles. For $V<V_{sw}$ the north pole is stable. For $V>V_{sw}$ the south pole is stable. $(b)$ In the stationary regime $I_l=I_r=I$. For $V \neq V_{sw}$ the conductance differs between the steady state solution  
in the north pole, and that in the south poles (cf. Eqs.~(\ref{eq:I00}) and (\ref{eq:I0pi})). For either case we obtain Ohm's law, i.e., straight lines (with different slopes) starting at the origin. The scenario shown is for $g_l^{\uparrow \uparrow} + g_l^{\downarrow \downarrow} > g_l^{\uparrow \downarrow} + g_l^{\downarrow \uparrow}$ which results in a higher conductance when the magnetization points at the north pole. In the opposite case, the conductance of the south pole is higher. The stability for voltage biased regime, however, is the same in both cases. 
In addition, we assume here $B>0$ and $\Gamma_{l}^{\downarrow} > \Gamma_{l}^{\uparrow}$, which results in $g_s<0$ and in $V_{sw} /B >0$. \label{costhetavsI}}
\end{figure*}

We now analyze the stability of the two stationary solutions with $\sin\theta_0 = 0$. 
We define $\theta= \theta_0 + \delta\theta$ and $V_d = V_{d0} + \delta V_d$ and 
eliminate the fast variable $\dot\phi$ using the first of Eqs.~\eqref{eom}. We observe, then, 
that to the lowest order in $\delta\theta$ and $\delta V_d$ the dynamics of $\delta \theta$ 
decouples from that of $\delta V_d$ and is governed by
\begin{equation}\label{eq:stability}
\delta \dot \theta = -  \frac{\cos \theta_0\, \alpha(\theta_0)}{1+\alpha^2(\theta_0)} \left[B + \frac{2 \Gamma_\Delta \Gamma_r}{\Gamma_\Sigma^2-\Gamma_\Delta^2} V \right] \delta \theta + O(\delta\theta \cdot\delta V_d)\ .
\end{equation}
The stability of the stationary solutions at the north pole ($\cos \theta_0 = 1$) and at the south pole ($\cos \theta_0 = -1$) is completely determined by the sign of $\left[ B + \frac{2 \Gamma_\Delta \Gamma_r}{\Gamma_\Sigma^2-\Gamma_\Delta^2} V\right]$. 
For example, if $B>0$ and $\Gamma_\Delta<0$ (the situation depicted in Fig.~\ref{costhetavsI}), this sign
is positive for $V< V_{sw}$ and negative for $V>V_{sw}$, where the switching voltage is given by
\begin{equation}
V_{sw} = - \frac{(\Gamma_\Sigma^2-\Gamma_\Delta^2)B}{2 \Gamma_r \Gamma_\Delta}\ .
\end{equation}
For $V< V_{sw}$, the north pole is stable while the south pole is unstable; whereas for $V > V_{sw}$, the situation reverses (cf. Fig.~\ref{costhetavsI}). 

In general, the stability analysis of the type of performed around Eq.~(\ref{eq:stability}) should be done with care.
Eq.~(\ref{eq:stability}) describes the relaxation of the fluctuations of a collective variable, $\theta$, under the assumption that the distribution function remains unchanged, i.e., it is still determined by $\theta_0$ and $V_{d0}$ (approximation of Eq.~(\ref{eq:kinetic0})). 
Should a collective variable deviate from the stationary value for a time longer than the relaxation time of the distribution function, the latter would adjust itself to the new conditions (we would have to take into account higher order terms in Eq.~(\ref{eq:kinetic expanded})). Note that, close to the switching voltage, $V_{sw}$, where $\left[ B + \frac{2 \Gamma_\Delta \Gamma_r}{\Gamma_\Sigma^2-\Gamma_\Delta^2} V\right]$ tends to 0, the relaxation of $\delta \theta$ becomes very slow. This can be viewed as a critical slowing down near a driven phase transition. Thus, we expect a richer dynamics close to the switching voltage.

\subsection{Role of the (quantum) zero mode} 
\label{subsec:RoleZeroMode}
So far, we ignored the zero mode $V_{d\, 0}^q$.  The importance of this mode was realized in Ref.~\cite{KamenevGefenZeroMode} and in subsequent papers~\cite{Sedlmayr2006,Nissan-Cohen2011}. We perform the variation of the action in Eq. \eqref{action expanded} with respect to $V_{d\, 0}^q$, which yields 
\begin{equation}
 C V_{d0}=-\frac{i}{2}\, \mathrm{tr} \left[ \tilde G_d^K \right] -  N_0\ . \label{zero mode eom}
\end{equation}
This equation determines the electrostatic potential on the dot, $V_{d0}=V_{d\, 0}^c$, fixed by the excess charge $-\frac{i}{2}\, \mathrm{tr} \left[ \tilde G_d^K \right] -  N_0$, and the capacity $C$. 
We are now, finally, able to determine the electrostatic potential of the dot $V_{d0}$. Indeed, the equations of motion (\ref{eom}) leave this quantity completely arbitrary. 
They contain only the combination $V_d - V_{d0}$, which vanishes once the stationary solution is achieved.  
The stationary value of $V_d = V_{d0}$ should, thus, be determined from Eq.~(\ref{zero mode eom}). Assuming the background charge, $N_0$, is such that at $V=0$ the total charge of the nano particle is zero, we obtain 
\begin{equation}
C V_{d0} = \sum_\sigma \rho_d^\sigma \int d\epsilon \frac{1}{2}\left[F(\epsilon) - F_d^\sigma(\epsilon)\right]\ ,
\end{equation}
where $F_d^\sigma(\epsilon)$ is given by Eq.~(\ref{Fd}) and depends on $V_{d0}$ itself. Solving this equation for $V_{d0}$ we obtain
\begin{eqnarray}
V_{d0} & = & \frac{1}{C+ \rho_d^\uparrow + \rho_d^\downarrow} \sum_\sigma \frac{\rho_d^\sigma}{\Gamma_\sigma(\theta_0)} \nonumber \\* & \times &\left[\left(\cos^2\frac{\theta_0}{2} \Gamma_l^\sigma +  \sin^2\frac{\theta_0}{2} \Gamma_l^{\bar\sigma} \right) V +  \Gamma_\Delta \sin^2 \theta_0 \frac{B_0}{2}\right]\ . \nonumber \\*
\end{eqnarray}
For the two stationary states this gives
\begin{equation}
V_{d0}(\theta_0=0)=\frac{V}{C+ \rho_d^\uparrow + \rho_d^\downarrow}\,\left(\frac{\rho_d^\uparrow\,\Gamma_l^\uparrow }{\Gamma_l^\uparrow + \Gamma_r} + \frac{\rho_d^\downarrow\,\Gamma_l^\downarrow }{\Gamma_l^\downarrow + \Gamma_r} \right)\ ,
\end{equation}
and
\begin{equation}
V_{d0}(\theta_0=\pi)=\frac{V}{C+ \rho_d^\uparrow + \rho_d^\downarrow}\,\left(\frac{\rho_d^\uparrow\,\Gamma_l^\downarrow }{\Gamma_l^\downarrow + \Gamma_r} + \frac{\rho_d^\downarrow\,\Gamma_l^\uparrow }{\Gamma_l^\uparrow + \Gamma_r} \right)\ .
\end{equation}

\section{Summary and discussion}
\label{sec:Discussion}

The setup studied here is paradigmatic in the field of spintronics. We have revisited the dynamics of magnetization of a single-domain ferromagnetic nano particle (quantum dot) under the influence of
spin-transfer torque (STT). Earlier studies have predicted persistent
precession of magnetization driven by STT. This effect is predicted even in ferromagnetic nano
particles with no actual anisotropy, i.e., the only preferred direction is due to the applied field. Analyzing the semiclassical magnetization dynamics of such a system we have found that this regime of persistent precession may become unstable owing to the non-equilibrium distribution of electrons on the dot (Eq.~\ref{Fd}).  Consequently, the persistent precession is replaced by a different steady state spin dynamics, leading to stable spin orientation either along the north pole or the south pole of the Bloch sphere (Eqs.~(\ref{eq:EqFortheta0}) and (\ref{eq:stability}); Fig. 2a). Each of these precession-less spin orientations is associated with different values of Ohmic conductance 
(Eqs.~(\ref{eq:I00}) and (\ref{eq:I0pi}); Fig. 2b). The traditional physics of spin precession may be restored  by further introducing internal energy and spin equilibration processes in the
quantum dot.

The effect we report here is rather subtle. One could pursue the following erroneous strategy. 
Assuming the right lead (sink) is coupled to the quantum dot much stronger than the left lead, 
$\Gamma_r \gg \Gamma^\sigma_l$, one could approximate in Eq.~(\ref{Fd})
\begin{eqnarray}
\label{Fd_approx}
F_d^\sigma(\epsilon) & \approx & \cos^2 \frac{\theta_0}{2}  F(\epsilon - \sigma B_- +V_{d0}) \nonumber \\*
    & + & \sin^2 \frac{\theta_0}{2} F(\epsilon - \bar \sigma B_+ +V_{d0}) \ ,
\end{eqnarray}
which would lead in (\ref{Vsigma}) to $V_{\sigma}(\theta_0) \approx V_{d0}$. In addition, one could argue (correctly) 
that in this regime almost the entire voltage drops on the left junction, implying $V_{d} = V_{d0} \ll V$. We would then 
obtain $I_s \approx g_s V$, and the second equation of (\ref{eom}) would read
\begin{equation}\label{2ndeomapprox}
\dot \theta  =  - \sin \theta \left[-\alpha(\theta) \dot \phi + (g_s/S) V \right] \ .
\end{equation}
This would lead to the persistent precession solution as in Ref.~\cite{ChudnovskiyPRL}. 

The reason why this seemingly plausible line of reasoning  does not work is the following. 
We recall that the Gilbert coefficient is given by $\alpha(\theta)= \frac{1}{S} (\tilde g_l(\theta) + \tilde g_r)$.
For $\Gamma_r \propto \tilde g_r \rightarrow \infty$ the Gilbert friction is large and one would need a very large voltage $V$ in (\ref{2ndeomapprox}) 
to set the r.h.s. of Eq.~(\ref{2ndeomapprox})  to zero (still keeping $\sin\theta$ finite). The angle dependent part of $\alpha(\theta)$, which is crucial for the emergence of the spin-precession solution with $\sin\theta_0 \neq 0$, is now a small correction on top of the main, $\theta$-independent, part determined by 
$\tilde g_r$. This correction should now be compared with the small non-equilibrium corrections to the spin-torque current 
related to the parts of the distribution function dropped in Eqs.~(\ref{Fd_approx}) (cf.~Eq.~(\ref{Fd})).
The reason why these parts of the distribution function $F_d^\sigma$ 
are important is as follows. The proper distribution function (\ref{Fd}) has the usual two-step structure 
around the electrochemical potentials of the left and the right leads (each step is in addition split to two by spin precession). In the regime $\Gamma_r \gg \Gamma^\sigma_l$ the step due to the left lead has a very small height.
Thus, it seems reasonable to neglect it. Yet, if the voltage $V$ is high, the weight of this step (integrated over energies) 
is not negligible. 
In Appendix~\ref{sec:EnforcedDistFunc} we discuss the case of enforced distribution function in the dot, in which 
case Eq.~(\ref{2ndeomapprox}) holds and one does get persistent precession.

We note in passing that the regime $\tilde g_r \gg \tilde g_l$ seems to be realized in real spin torque nano-oscillators~\cite{Zeng2013,Choi2014,Ramaswamy2016}. Indeed, in 
Refs.~\cite{Zeng2013,Choi2014,Ramaswamy2016} the coupling of the free ferromagnetic 
layer (nano particle) to the fixed ferromagnet is via a tunnel junction, whereas the contact with the (non-ferromagnetic) sink is the direct one. At the same time, our model disregards the strong anisotropy, crucial in real systems, as well as the internal dissipation within the nano particle. Further studies are then necessary to establish the applicability of our results to systems like those investigated in Refs.~\cite{Zeng2013,Choi2014,Ramaswamy2016}.

Finally, let us put our analysis in the general framework of charge and spin transport, referring to the analysis of Sec.~\ref{subsec:RoleZeroMode}. The stationary distribution function (\ref{Fd}), and the parameters controlling it, i.e., $\theta_0$, $V_{d0}$, $B_0$, are determined, {\it inter alia}, by the stationary Kirchhoff's condition $I_l = I_r$. Had we have full equilibration in the dot, this condition would define the {\it electrochemical} potential of the dot. This procedure  is not restricted to equilibrium conditions. Under non-equilibrium conditions it is possible to  define a (spin resolved) electrochemical potential. This is done by weakly coupling  (as a gedanken experiment) the dot to thermodynamic  spin-polarized reservoirs, and adjusting their respective electrochemical potentials such that no current flows from/to these reservoirs. By contrast, the {\it electrostatic}
potential of the dot, $V_d$, is not defined by  Kirchhoff's condition. Rather it is defined by the requirement that the charge of the dot (determined by the distribution function) is equal to $C V_d$. Clearly, the value of the {\it chemical} potential should be determined to be consistent with the electrochemical and the electrostatic potentials of the dot.

\section{Acknowledgements}
We thank A. Altland and A. Kamenev for fruitful discussions. 
We acknowledge support from DFG Research Grant SH 81/3-1,
ISF grant 1349/14, CRC 183 of the DFG, the IMOS Israel-Russia program,
the Minerva Foundation, Russian Foundation for Basic Research (Grant No. 15-52-06005), 
Ministry of Education and Science of the Russian Federation (Grant No. MD-5620.2016.2), and Russian President Scientific Schools (Grant NSh-10129.2016.2).

\appendix

\section{Off-diagonal elements of the self-energy}
\label{sec:AppOffDiag}

\subsection{Self-Energies and relaxation rates}

The self-energies due to the left lead are given by
\begin{eqnarray}
\Sigma^{R/A}_{l,\alpha \beta,\sigma}(\epsilon) &=& \sum_{n=1}^{N_l} \int \frac{dk}{2\pi} \frac{t^{\phantom *}_{\alpha, n}\, t_{\beta, n}^*}{\epsilon - \epsilon_{n k} + \frac{M_\mathrm{fix}}{2}\sigma -V \pm i 0 } \nonumber \ , \\
\Sigma^K_{l,\alpha \beta,\sigma}(\epsilon) &=& - 2\pi i  \sum_{n=1}^{N_l} \int \frac{dk}{2\pi} t^{\phantom *}_{\alpha, n}\, t_{\beta, n}^* \nonumber \\* & \times & \delta\left(\epsilon -  \epsilon_{n k}+ \frac{M_\mathrm{fix}}{2}\sigma -V\right) \,F_l(\epsilon)\ ,
\end{eqnarray}
where $F_l(\epsilon)\equiv 1-2 n_l(\epsilon)$.
We remind that the left lead distribution function reads $n_l(\epsilon)=1/(e^{\beta (\epsilon - (\mu +V))}+1)$, 
where $\mu$ is the bare chemical potential. Neglecting the principal value parts gives
\begin{eqnarray}
\Sigma^{R/A}_{l,\alpha \beta,\sigma}(\epsilon) & \approx & \mp i \Gamma^{\sigma}_{l,\alpha\beta}(\epsilon)\ , \nonumber \\
\Sigma^{K}_{l,\alpha \beta,\sigma}(\epsilon) & \approx & -2i \Gamma^{\sigma}_{l,\alpha\beta}(\epsilon) \,F_l(\epsilon)\ ,
\end{eqnarray}
where $\Gamma^{\sigma}_{l,\alpha\beta}(\epsilon) =\pi \sum_{n=1}^{N_l} \rho_n\left(\epsilon+\frac{M_\mathrm{fix}}{2}\sigma-V\right) \, t^{\phantom *}_{\alpha, n} t^{*}_{\beta,n}$. Due to the large value of $M_\mathrm{fix}$ the densities 
of states for spin up and down are different. Thus, the matrices $\Gamma^{\sigma}_{l,\alpha\beta}$ are spin-dependent.

For the self-energies due to the right lead we obtain
\begin{eqnarray}
\Sigma^{R/A}_{r,\alpha \beta,\sigma}(\epsilon) &=& \sum_{n=N_l + 1}^{N_l + N_r} \int \frac{dk}{2\pi} \frac{t^{\phantom *}_{\alpha, n}\, t_{\beta, n}^*}{\epsilon - \epsilon_{n k} \pm i 0 }\nonumber \ ,\\
\Sigma^K_{r,\alpha \beta,\sigma}(\epsilon) &=& - 2\pi i  \sum_{n=N_l + 1}^{N_l + N_r} \int \frac{dk}{2\pi}\, t^{\phantom *}_{\alpha, n}\, t_{\beta, n}^* \nonumber \\* & \times & \delta(\epsilon -  \epsilon_{n k}) \,F_r(\epsilon)\ , 
\end{eqnarray}
where $F_r(\epsilon)\equiv 1-2 n_r(\epsilon)$ and $n_r(\epsilon)=1/(e^{\beta (\epsilon - \mu)}+1)$.
Neglecting the principal value contributions we obtain
\begin{eqnarray}
\Sigma^{R/A}_{r,\alpha \beta,\sigma}(\epsilon) & \approx & \mp i \Gamma_{r,\alpha\beta}(\epsilon) \nonumber \\
\Sigma^{K}_{r,\alpha \beta,\sigma}(\epsilon) & \approx & -2i \Gamma_{r,\alpha\beta}(\epsilon)\, F_r(\epsilon)\ .
\end{eqnarray}
where $\Gamma_{r,\alpha\beta}(\epsilon) =\pi \sum_{n=N_l + 1}^{N_l + N_r} \rho^{\phantom *}_n(\epsilon) \, t^{\phantom *}_{\alpha, n} t^{*}_{\beta,n}$. As the right lead is non-magnetic, the rates $\Gamma_{r,\alpha\beta}(\epsilon)$ are spin-independent.
We assume that the densities of states $\rho_n$ do not depend strongly on energy on the scale 
of $V$ or $B$. Thus we can assume that $\Gamma^{\sigma}_{l,\alpha\beta}$ and $\Gamma_{r,\alpha\beta}$ are independent of $\epsilon$
for $| \epsilon | \lesssim |V|, |B|$.

\subsection{Reason for neglecting the off-diagonal elements}
Here we consider only one lead, e.g., the right one. The self-energy of the lead reads
\begin{eqnarray}
\Sigma_{\alpha \beta}(\epsilon) & = & \sum_n \int \frac{dk}{2\pi} \frac{t^{\phantom *}_{\alpha, n}\, t_{\beta, n}^*}{\epsilon - \epsilon_{n k} +i 0 }\nonumber \\
& \approx & - i \pi \sum_n \underbrace{\int \frac{dk}{2\pi}\, \delta(\epsilon - \epsilon_{n k})}_{= \rho_n(\epsilon)} t^{\phantom *}_{\alpha, n}\, t_{\beta, n}^*\nonumber \\
& = &- i \sum_n \Gamma_{n, \alpha \beta} (\epsilon)\ ,
\end{eqnarray}
where in the first step, we neglected the principal part, and we defined $\Gamma_{n, \alpha \beta} ( \epsilon) = \pi \rho_n ( \epsilon ) t^{\phantom *}_{\alpha, n}\, t_{\beta, n}^*$\ , where $\rho_n ( \epsilon)$ is the density of states for the $n$-th channel.

We assume that $t_{\alpha, n}$ are random. This randomness will affect $\Gamma_{n, \alpha \beta}$. Note that, $\Gamma_{n, \alpha \alpha}>0$, whereas $\Gamma_{n, \alpha \beta}$ may be complex/negative for $\alpha \neq \beta$. We define the total broadening $\Gamma_{\alpha \beta}$ as a sum over all channels, i.e.
\begin{equation}
\Gamma_{\alpha \beta}= \sum_n \Gamma_{n, \alpha \beta}\ .
\end{equation}
Since $\Gamma_{n, \alpha \alpha}>0$, we expect
\begin{equation}
\Gamma_{\alpha \alpha} = \sum_n \Gamma_{n, \alpha \alpha} \approx \gamma_1\, N\ ,
\end{equation}
whereas for $\alpha \neq \beta$, we expect
\begin{equation}
\Gamma_{\alpha \beta} = \sum_n \Gamma_{n, \alpha \beta} \approx \gamma_1\, \sqrt{N}\ ,
\end{equation}
where $\gamma_1$ is the average broadening per channel and $N$ is the number of channels. For the tunneling regime, we assume $\gamma_1 \approx \pi |t|^2 \rho_1 \ll \delta_d$, where $\rho_1$ is the average density of states per channel and $\delta_d$ is the mean level spacing in the dot.

Now, we consider the consequence for the Green's function and for the AES 
kernel $\alpha$ (see Eq.~(\ref{eq:SAESR})). In what follows we disregard the Keldysh structure, as 
it is not important for the estimates. We obtain
\begin{equation}
G_{\alpha \beta}^{-1}= \underbrace{\left( \epsilon - \epsilon_\alpha - \Sigma_{\alpha \alpha} \right)}_{=: G_{0, \alpha}^{-1}} \delta_{\alpha \beta} - \delta \Sigma_{\alpha \beta}\ , 
\end{equation}
where we defined $\delta \Sigma_{\alpha \beta} = \Sigma_{\alpha \beta} (1 - \delta_{\alpha \beta})$ . 
From $G^{-1} G = \mathbf 1$
we obtain
\begin{eqnarray}\label{eq:Gexpansion}
G_{\alpha \beta} & = &  G_{0, \alpha}\, \delta_{\alpha \beta}  + G_{0, \alpha}\,\delta\Sigma_{\alpha \beta}\, G_{0, \beta} \nonumber \\* & + & \sum_\gamma G_{0, \alpha} \,\delta \Sigma_{\alpha \gamma}\,G_{0, \gamma}\,\delta \Sigma_{\gamma \beta}\,G_{0, \beta}  + \dots \ .
\end{eqnarray}
For the AES kernel we obtain (the Keldysh and spin structures are omitted for clarity)
\begin{equation}\label{eq:alphaexpansion}
\alpha(\epsilon) \sim \sum_\alpha G_{0,\alpha}\Sigma_{\alpha\alpha} + \sum_{\alpha\neq\beta}G_{0, \alpha}\,\delta\Sigma_{\alpha \beta}\, G_{0, \beta} \,\delta\Sigma_{\beta \alpha} + \dots.
\end{equation}
Note that, $G_{0, \alpha}= (\epsilon - \epsilon_\alpha + i \Gamma_{\alpha \alpha})^{-1}$ and at the resonance $G_{0, \alpha} \propto (N \gamma_1)^{-1}$, whereas $\delta \Sigma_{\alpha\alpha} \sim \Gamma_{\alpha\alpha} \propto N \gamma_1$ and $\delta \Sigma_{\alpha\beta} \sim \Gamma_{\alpha\beta}\big|_{\alpha \neq \beta} \propto \sqrt{N} \gamma_1$ (here we used the retarded $G_{0, \alpha}$, but the argument holds for all Keldysh components). The leading term of (\ref{eq:alphaexpansion}) is, thus, of order 
$(N\gamma_1)^{-1} (N\gamma_1) (N\gamma_1/\delta_d) \sim (N\gamma_1/\delta_d) \sim g$, where $g$ is dimensionless conductance. Here the last multiplier 
$N\gamma_1/\delta$ is the number of resonant contributions. The next term in the expansion (\ref{eq:alphaexpansion})
is of order $(N\gamma_1)^{-2} (\sqrt{N}\gamma_1)^2 (N\gamma_1/\delta_d)^2 \sim g^2/N$ (there are $(N\gamma_1/\delta_d)^2$ resonant contributions here). The tunneling 
approximation is justified if $g/N \sim \gamma_1/\delta_d \ll 1$. The higher order terms in the expansion (\ref{eq:alphaexpansion}) can be analyzed similarly. Thus, in the limit $N\gg 1$ we can restrict ourselves 
to the leading term in (\ref{eq:alphaexpansion}).

We conclude that in the limit of large number of weakly coupled (tunneling) transverse channels it is allowed to replace the matrices $\Gamma^{\sigma}_{l,\alpha,\beta}$ and $\Gamma_{r,\alpha,\beta}$ by diagonal $\epsilon$-independent
matrices, i.e., $\Gamma^{\sigma}_{l,\alpha,\beta} \rightarrow \Gamma^{\sigma}_{l} \delta_{\alpha,\beta}$ and 
$\Gamma_{r,\alpha,\beta} \rightarrow \Gamma_{r} \delta_{\alpha,\beta}$. Thus the system is described by just 
three rates, $\Gamma^{\uparrow}_{l}$, $\Gamma^{\downarrow}_{l}$, and $\Gamma_{r}$.

\section{Enforced distribution function}
\label{sec:EnforcedDistFunc}

To relate our results to those in the literature, we enforce an equilibrium distribution function on the dot in the rotating frame, $F_d^\sigma(\epsilon)=F(\epsilon)$. In this case we obtain
$I_s = g_s (V-V_d)$,  $I_l = 4 g_l (V-V_d)  - g_s \sin^2 \theta \, \dot \phi$, 
$I_r = 4 g_r V_{d}$. 
This gives the following equations of motion
\begin{eqnarray}
\sin\theta\,\dot \phi & = & - \sin \theta B - \alpha(\theta)\, \dot \theta\ , \nonumber \\ 
\sin\theta\,\dot \theta & = &  \sin^2 \theta\, \left[ \alpha(\theta)\dot \phi - g_s(V-V_d)/S \right]\ , \nonumber \\ 
C \dot V_d &=& 4g_l (V-V_d)  - g_s \sin^2 \theta \, \dot \phi - 4 g_r V_d\ . \label{eom2}
\end{eqnarray}
The second equation would again predict a persistent precession state (cf.~\cite{ChudnovskiyPRL}), if we were able to control the voltage on the left junction $V_l \equiv V - V_d$. 

The distribution function enforced here could emerge due to strong energy and spin relaxation. In this case, however, one could expect an extra internal contribution to Gilbert damping coefficient.

\section{Calculation of $iS_{WZNW}$\label{sec:calc.WZNW}}

Here, we calculate the Berry-phase (WZNW) action. We start from,
\begin{equation}
i \mathcal{S}_{WZNW}= - \frac{1}{2} \int dt\, \mathrm{tr} [ \tilde G_d^K(t,t) Q^q(t)]\ . 
\end{equation}
We use $Q^q = \left( \dot \phi^c p^q - \phi^q \dot p^c \right) \frac{\sigma_z}{2}$ and take the trace over spin space to obtain
\begin{eqnarray}
i \mathcal{S}_{WZNW}= & - & \frac{1}{4} \int dt\, \left( \dot \phi^c p^q - \phi^q \dot p^c \right) \nonumber \\* & \times & \mathrm{tr} \left[ \tilde G_{d \uparrow}^K(t,t) -\tilde G_{d \downarrow}^K(t,t) \right] \ .
\end{eqnarray}
With $\tilde G_{d \sigma}^K(t,t) = \int \frac{d\epsilon}{2 \pi} \tilde G_{d \sigma}^K(\epsilon) = - i \int d\epsilon\, \frac{1}{\pi} \frac{\Gamma_\sigma (\theta)}{(\epsilon - \epsilon_\alpha + \frac{M_0}{2} \sigma)^2 + \Gamma_\sigma (\theta)^2} F_{d}^\sigma(\epsilon)$ we obtain
\begin{widetext}
\begin{eqnarray}\label{eq:calculatingWZNW}
\frac{1}{2}\, \mathrm{tr} \left[ \tilde G_{d \uparrow}^K(t,t) -\tilde G_{d \downarrow}^K(t,t) \right] & = &  \frac{- i}{2} \int d\epsilon\, \left[ \rho_d \left( \epsilon + \frac{M_0}{2} , \Gamma_\uparrow \right) F_d^\uparrow(\epsilon) -  \rho_d \left( \epsilon - \frac{M_0}{2} , \Gamma_\downarrow \right) F_d^\downarrow(\epsilon) \right] \nonumber \\*
& \approx &  \frac{- i}{2} \int d\epsilon\, \rho_d(\epsilon) \left[ F_d^\uparrow\left(\epsilon - \frac{M_0}{2} \right)- F_d^\uparrow\left(\epsilon + \frac{M_0}{2}\right) \right] =:   2 i S\ ,
\end{eqnarray}
\end{widetext}
where we define the coarse-grained density of states $\rho_d \left( \epsilon, \Gamma_\sigma \right) \equiv \sum_\alpha \frac{1}{\pi} \frac{\Gamma_\sigma(\theta)}{ \left( \epsilon - \epsilon_\alpha \right)^2 + \Gamma_\sigma(\theta)^2}$. We approximate $\rho_d(\epsilon, \Gamma_\uparrow) \approx \rho_d (\epsilon, \Gamma_\downarrow) =: \rho_d(\epsilon)$ in the second line of (\ref{eq:calculatingWZNW})(the difference is only in the broadening $\Gamma_\sigma$). As defined in (\ref{eq:calculatingWZNW}), the total spin, $S$, is about half the number of states in the interval $[\mu- \frac{M_0}{2} , \mu + \frac{M_0}{2}]$. Putting everything together, we obtain Eq. \eqref{eq:WZNW}.

\section{Calculation of $iS_{AES}$\label{sec:calc.AES}}

Here, we provide details of the calculation of the integral kernels $\alpha_{\sigma \sigma'}^R(\epsilon)$ and
$\alpha_{\sigma \sigma'}^K(\epsilon)$, which appear in $iS_{AES}$ (see Eqs.~(\ref{eq:SAESR},\ref{eq:SAESK})).
For the retarded kernel we obtain
\begin{eqnarray}
&&\alpha_{\sigma \sigma'}^R(\epsilon)  = \nonumber\\&& =\int \frac{d \epsilon'}{2 \pi} \mathrm{tr} \left[ G_{d \sigma}^K(\epsilon') \Sigma_{\sigma'}^A(\epsilon' -\epsilon) + G_{d \sigma}^R(\epsilon') \Sigma_{\sigma'}^K(\epsilon' - \epsilon) \right]\nonumber \\
&& =  \int \frac{d \epsilon'}{2 \pi} \sum_\alpha \left[ G_{d \alpha \sigma}^K(\epsilon') \Sigma_{ \alpha \alpha \sigma'}^A(\epsilon' -\epsilon)\right.\nonumber\\  &&+ \left. G_{d \alpha \sigma}^R(\epsilon') \Sigma_{ \alpha \alpha \sigma'}^K(\epsilon' - \epsilon) \right]\ .
\end{eqnarray}
Drawing on the discussion in appendix \ref{sec:AppOffDiag}, we approximate $\Sigma_{ \alpha \alpha \sigma'}^A(\epsilon' -\epsilon) \approx  i (\Gamma_l^{\sigma'} + \Gamma_r )$ and $\Sigma_{ \alpha \alpha \sigma'}^K(\epsilon' -\epsilon) \approx - 2 i [\Gamma_l^{\sigma'} F_l ( \epsilon' - \epsilon ) + \Gamma_r  F_r ( \epsilon' - \epsilon)]$ and obtain
\begin{eqnarray}
\alpha_{\sigma \sigma'}^R(\epsilon)  &=& \int \frac{d \epsilon'}{2\pi} \sum_\alpha  \Bigg[  \frac{2(\Gamma_l^{\sigma'} + \Gamma_r)\Gamma_\sigma(\theta_0)F_d^\sigma(\epsilon') }{(\epsilon' - \epsilon_\alpha + \frac{M_0}{2} \sigma)^2 + \Gamma_\sigma^2(\theta_0)}   \nonumber \\ & - &  \frac{2i \left( \Gamma_l^{\sigma'} F_l(\epsilon' -\epsilon) + \Gamma_r F_r( \epsilon' - \epsilon) \right)}{\epsilon' - \epsilon_\alpha + \frac{M_0}{2} \sigma + i \Gamma_\sigma(\theta_0) }   \Bigg]\ .
\end{eqnarray}
Now, we calculate the real part and obtain
\begin{eqnarray}
\mathrm{Re} \left[ \alpha_{\sigma \sigma'}^R(\epsilon) \right] &=& \underbrace{2 \rho_d^\sigma \Gamma_l^{\sigma'}}_{g_l^{\sigma \sigma'}} \, \frac{1}{2} \int d\epsilon' \left( F_d^\sigma(\epsilon') - F_l ( \epsilon' - \epsilon) \right) 
\nonumber\\ &+& \underbrace{2 \rho_d^\sigma \Gamma_r}_{g_r^{\sigma \sigma'}}\, \frac{1}{2} \int d\epsilon' \left( F_d^\sigma(\epsilon' ) - F_r(\epsilon' - \epsilon) \right)\ ,\nonumber\\
\end{eqnarray}
where $\rho_d^\sigma (\epsilon) \equiv \rho_d\left(\epsilon + \frac{M_0}{2}\sigma\right)$ and we assume that 
$\rho_d^\sigma (\epsilon)$ is approximately $\epsilon$-independent on the scale of $|\epsilon| \ll M_0$. Analogous to the calculation of the WZNW action (appendix \ref{sec:calc.WZNW}), we disregard small correction to the density of states arising from the difference in $\Gamma_\uparrow(\theta_0)$ and $\Gamma_\downarrow(\theta_0)$. The connection to equations \eqref{alphal} and \eqref{alphar} is now straightforward.

The Keldysh component,
\begin{eqnarray}
&&\alpha_{\sigma \sigma'}^K(\epsilon) =  \int \frac{d \epsilon'}{2 \pi} \mathrm{tr} \left[ G_{d \sigma}^K(\epsilon') \Sigma_{\sigma'}^K(\epsilon' -\epsilon)\right. \nonumber \\ &&+ \left. G_{d \sigma}^R(\epsilon') \Sigma_{\sigma'}^A(\epsilon' - \epsilon) + G_{d \sigma}^A(\epsilon') \Sigma_{\sigma'}^R(\epsilon' - \epsilon) \right] \ ,
\end{eqnarray}
can be calculated similarly.  We obtain  $\alpha^{K}_{\sigma \sigma'}(\epsilon)= g^{\sigma \sigma'}_l \alpha^K_{l,\sigma}(\epsilon)+ g^{\sigma \sigma'}_r \alpha^K_{r,\sigma}(\epsilon)$  with
\begin{widetext}
\begin{eqnarray}
\alpha^K_{l,\sigma} (\epsilon)  = && - \int d\epsilon' \left[ F_d^\sigma(\epsilon') F(\epsilon'-\epsilon - V) -1 \right] 
\nonumber \\
 = &&  \frac{\Gamma_l^\sigma  \cos^2 \frac{\theta_0}{2}}{\Gamma_\sigma (\theta_0)}\ 2 (\epsilon - \sigma B_- + V_{d0})\, \coth \frac{\epsilon - \sigma B_- + V_{d0}}{2T}  \nonumber \\
& & + \frac{\Gamma_l^{\bar \sigma}  \sin^2 \frac{\theta_0}{2}}{\Gamma_\sigma (\theta_0)}\ 2 (\epsilon - \bar \sigma B_+ + V_{d0})\, \coth \frac{\epsilon - \bar \sigma B_+ + V_{d0}}{2T} \nonumber \\
& & + \frac{\Gamma_r  \cos^2 \frac{\theta_0}{2}}{\Gamma_\sigma (\theta_0)}\ 2 (\epsilon - \sigma B_- + V_{d0} + V)\, \coth \frac{\epsilon - \sigma B_- + V_{d0} + V}{2T} \nonumber \\
& & + \frac{\Gamma_r  \sin^2 \frac{\theta_0}{2}}{\Gamma_\sigma (\theta_0)}\ 2 (\epsilon - \bar \sigma B_+ + V_{d0} +V)\, \coth \frac{\epsilon - \bar \sigma B_+ + V_{d0}+V}{2T} \ ,
\end{eqnarray}
and
\begin{eqnarray}
\alpha^K_{r,\sigma} (\epsilon)  = && - \int d\epsilon' \left[ F_d^\sigma(\epsilon') F(\epsilon'-\epsilon ) -1 \right] \nonumber \\
=  &&  \frac{\Gamma_l^\sigma  \cos^2 \frac{\theta_0}{2}}{\Gamma_\sigma (\theta_0)}\ 2 (\epsilon - \sigma B_- + V_{d0}-V)\, \coth \frac{\epsilon - \sigma B_- + V_{d0}-V}{2T}  \nonumber \\
& & + \frac{\Gamma_l^{\bar \sigma}  \sin^2 \frac{\theta_0}{2}}{\Gamma_\sigma (\theta_0)}\ 2 (\epsilon - \bar \sigma B_+ + V_{d0}-V)\, \coth \frac{\epsilon - \bar \sigma B_+ + V_{d0}-V}{2T} \nonumber \\
& & + \frac{\Gamma_r  \cos^2 \frac{\theta_0}{2}}{\Gamma_\sigma (\theta_0)}\ 2 (\epsilon - \sigma B_- + V_{d0} )\, \coth \frac{\epsilon - \sigma B_- + V_{d0} }{2T} \nonumber \\
& & + \frac{\Gamma_r  \sin^2 \frac{\theta_0}{2}}{\Gamma_\sigma (\theta_0)}\ 2 (\epsilon - \bar \sigma B_+ + V_{d0} )\, \coth \frac{\epsilon - \bar \sigma B_+ + V_{d0}}{2T}\ . 
\end{eqnarray}
\end{widetext}
Using $\alpha^{K}_{\sigma \sigma'}(\epsilon)$ one could calculate, e.g., the statistics of spin and charge currents similarly to Ref.~\cite{VirtanenHeikkila2016}.

\bibliography{paper.bib}

\end{document}